\documentclass[aps,prb,twocolumn,10pt,superscriptaddress]{revtex4-1}

\usepackage{graphicx}
\usepackage{tabularx}
\usepackage{booktabs}
\usepackage{hyperref}
\usepackage[all]{hypcap}
\usepackage{natmove} 

\usepackage[version=4]{mhchem}
\usepackage{siunitx} 

\hypersetup{
	pdfnewwindow=true,      
    colorlinks=true,       
    linkcolor=blue,          
    citecolor=blue,        
    filecolor=blue,      
    urlcolor=blue           
}

\begin{document}

\title{Excitonic Absorption Signatures of Twisted Bilayer WSe$_{2}$ by Electron Energy-Loss Spectroscopy}

\author{Steffi Y. Woo}
\email[]{steffi.woo@universite-paris-saclay.fr}
\affiliation{Laboratoire de Physique des Solides, Universit\'{e} Paris-Saclay, CNRS UMR 8502, F-91405, Orsay, France}

\author{Alberto Zobelli}
\affiliation{Laboratoire de Physique des Solides, Universit\'{e} Paris-Saclay, CNRS UMR 8502, F-91405, Orsay, France}

\author{Robert Schneider}
\affiliation{Institute of Physics and Center for Nanotechnology, University of M\"{u}nster, M\"{u}nster, Germany}

\author{Ashish Arora}
\affiliation{Institute of Physics and Center for Nanotechnology, University of M\"{u}nster, M\"{u}nster, Germany}
\affiliation{Indian Institute of Science Education and Research, Dr. Homi Bhabha Road, 411008 Pune, India}

\author{Johann A. Preu\ss}
\affiliation{Institute of Physics and Center for Nanotechnology, University of M\"{u}nster, M\"{u}nster, Germany}

\author{Benjamin J. Carey}
\affiliation{Institute of Physics and Center for Nanotechnology, University of M\"{u}nster, M\"{u}nster, Germany}

\author{Steffen Michaelis de Vasconcellos}
\affiliation{Institute of Physics and Center for Nanotechnology, University of M\"{u}nster, M\"{u}nster, Germany}

\author{Maurizia Palummo}
\affiliation{Dipartimento di Fisica, and European Theoretical Spectroscopy Facility (ETSF), Universit\`{a} di Roma Tor Vergata, Via della Ricerca Scientifica 1, 00133 Rome, Italy}

\author{Rudolf Bratschitsch}
\affiliation{Institute of Physics and Center for Nanotechnology, University of M\"{u}nster, M\"{u}nster, Germany}

\author{Luiz H. G. Tizei}
\email[]{luiz.galvao-tizei@universite-paris-saclay.fr}
\affiliation{Laboratoire de Physique des Solides, Universit\'{e} Paris-Saclay, CNRS UMR 8502, F-91405, Orsay, France}

\begin{abstract}
Moir\'{e} twist angle underpins the interlayer interaction of excitons in twisted van der Waals hetero- and homo-structures. The influence of twist angle on the excitonic absorption of twisted bilayer tungsten diselenide (WSe$_{2}$) has been investigated using electron energy-loss spectroscopy. Atomic-resolution imaging by scanning transmission electron microscopy was used to determine key structural parameters, including the nanoscale measurement of the relative twist angle and stacking order. Detailed spectral analysis revealed a pronounced blueshift in the high-energy excitonic peak C with increasing twist angle, up to 200 meV when compared to the AA$^{\prime}$ stacking. The experimental findings have been discussed relative to first-principle calculations of the dielectric response of the AA$^{\prime}$ stacked bilayer WSe$_{2}$ as compared to monolayer WSe$_{2}$ by employing the \textit{GW} plus Bethe-Salpeter equation (BSE) approaches, resolving the origin of higher energy spectral features from ensembles of excitonic transitions, and thus any discrepancies between previous calculations. Furthermore, the electronic structure of moir\'{e} supercells spanning twist angles of $\sim$9.5--46.5$^{\circ}$ calculated by density functional theory (DFT) were unfolded, showing an uplifting of the conduction band minimum near the $Q$ point and minimal change in the upper valence band concurrently. The combined experiment/theory investigation provides valuable insight into the physical origins of high-energy absorption resonances in twisted bilayers, which enables to track the evolution of interlayer coupling from tuning of the exciton C transitions by absorption spectroscopy.
\end{abstract}


\maketitle

\section{Introduction}
Semiconducting transition metal dichalcogenides (TMDCs), such as tungsten diselenide (WSe$_{2}$), belong to a family of two-dimensional (2D) materials that exhibit fascinating electronic and optical properties. Due to interlayer coupling, bulk and multi-layered molybdenum- and tungsten-based TMDCs are indirect gap semiconductors, while their monolayers exhibit a crossover to a direct band gap. The strong spin-orbit interactions in TMDCs lead to spin-splitting of few hundred meV in the valence band and few to tens of meV in the conduction band, making TMDCs of interest for potential valleytronics applications.\cite{Wang2018review} The promising versatility of engineering with TMDCs is rooted in the flexibility of being artificially fabricated into van der Waals homo- or heterostructures. With the recent interest in \textit{magic-angle} twisted bilayer graphene for the appearance of flat bands \cite{Cao2018}, moir\'{e} superlattices of TMDCs naturally become a viable candidate for seeking similar phenomena \cite{Jin2019nat}. To date, in twisted homobilayer WSe$_{2}$ alone, evidence of low-energy flat bands \cite{Zhang2020flatbands} that could support emergent electronic phases for a continuum of low twist angles, such as superconductivity \cite{An2020} and correlated insulator states \cite{Wang2020} have been reported. Modification of the layer stacking as a function of twist alters the local atomic registry. As such, the interlayer coupling strength in homostructures has demonstrated to also be sensitive to twist angle in the case of bilayer MoS$_{2}$ \cite{Castellanos-Gomez2014} and WS$_{2}$ \cite{Yan2019}. Recent reports on domain formation caused by atomic reconstruction as a result of interlayer interactions of TMDC moir\'{e} superlattices with very low (near 0$^{\circ}$) twist angles in bilayer MoSe$_{2}$\cite{Sung2020}, WS$_{2}$ and WSe$_{2}$\cite{Weston2020} and double bilayer WSe$_{2}$\cite{An2020} have opened yet another avenue towards novel electronic and excitonic properties.

A major obstacle in the field involves correlating the optical response with imaging of the local structure of a nanometric moir\'{e} lattice, where efforts have turned towards various electron microscopies and scanning probe microscopies \cite{Andersen2021,McGilly2020,Weston2020}. The latest generation of electron monochromators with energy resolution of 10 meV and below has enabled the measurement of phonons in hexagonal boron nitride (\textit{h}-BN) and excitonic absorption in various monolayer TMDCs \cite{Tizei2015} in a transmission electron microscope with subwavelength spatial resolution. Aside from offering an exceptional combination of spatial and energy resolution, the high-energy incident electrons in electron energy-loss spectroscopy (EELS) have facile access to high-energy excitations of a few eV and beyond, unlike limitations met by optical techniques.

In this work, the combined high spatial and spectral resolution of aberration-corrected scanning transmission electron microscopy (STEM) and monochromated EELS in the low-loss regime were used to investigate the optical excitation response of atomically-thin WSe$_{2}$, specifically in twisted bilayer WSe$_{2}$ covering a large range of moir\'{e} angles. Relevant characteristics of the local atomic structure were also obtained within the same platform, including the twist angle and layer stacking. Furthermore, first-principle calculations of the electronic structure modifications in twisted bilayers and expected optical response relative to monolayers and zero-twist bilayers were used to interpret the changes in high-energy spectral features measured in EELS.

\section{Results}
\begin{figure}[tb]
	\centering
	\includegraphics[width=\columnwidth]{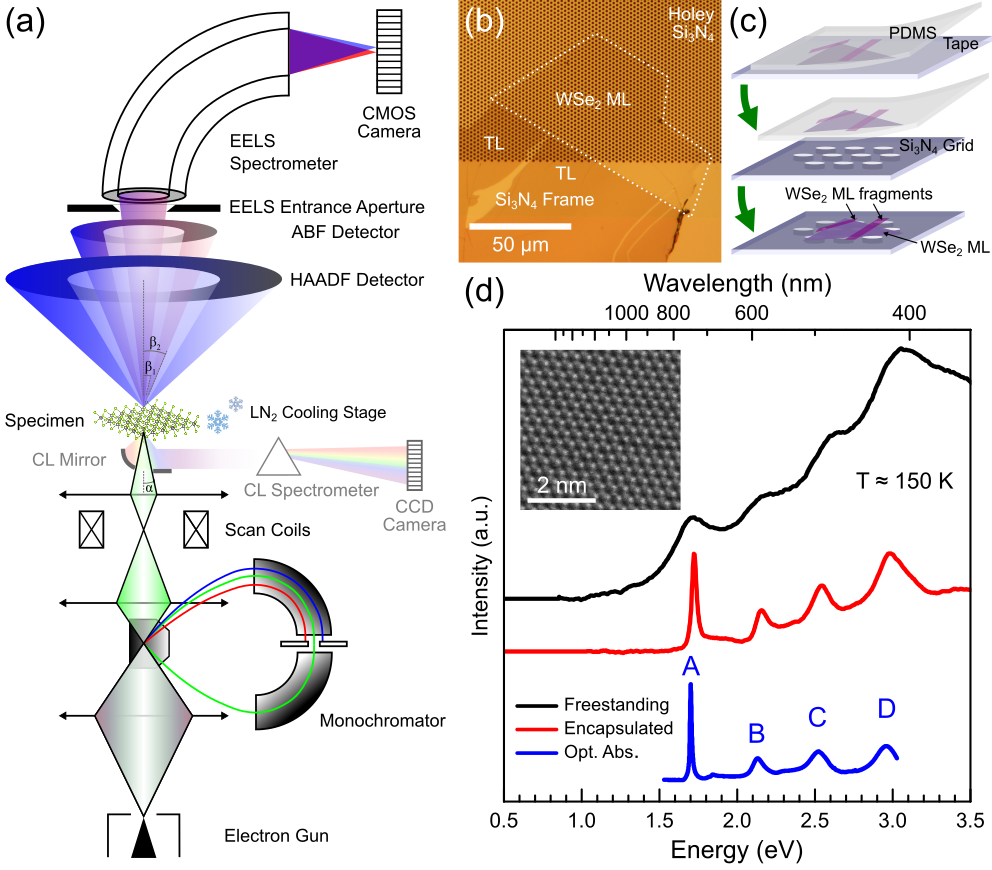}
	\caption{(a) Schematic of the electron microscope set-up including an electron monochromator, liquid nitrogen cooling sample holder, and a light collection system (not used in this work). (b) Optical microscopy image of a mechanically-exfoliated WSe$_{2}$ flake transferred onto a holey Si$_{3}$N$_{4}$ TEM grid. (c) Schematic illustrating how nanometric fragments of (twisted) bilayers and trilayers were likely formed during the exfoliation and transfer process. (d) Comparison of low-loss EELS spectra measured from freestanding (black) and \textit{h}-BN encapsulated (red) WSe$_{2}$ monolayer, and optical absorption spectrum of \textit{h}-BN encapsulated WSe$_{2}$ monolayer (blue), inset with a STEM-HAADF images of WSe$_{2}$ monolayer illustrating distinct contrast between the W and Se$_{2}$ atomic columns. \label{monolayer_opt}}
\end{figure}

Atomically-thin WSe$_{2}$ flakes have been mechanically exfoliated from a bulk synthetically-grown crystal \cite{Tonndorf2013}, and transferred onto a carbon-coated Si$_{3}$N$_{4}$ TEM grid with periodic 1 $\mu$m-sized holes. An optical microscopy reflectance image of exfoliated WSe$_{2}$ transferred onto the holey Si$_{3}$N$_{4}$ TEM grids with regions of different layer thickness is presented in Fig. \ref{monolayer_opt}(b), including the monolayer (ML) area marked by a dashed line and the adjacent trilayer (TL) regions. Atomically-resolved imaging has been performed on an aberration-corrected Nion UltraSTEM200 operated at 60 keV and monochromated EELS was performed on a modified Nion HERMES-S200 (also known as ChromaTEM) operated at 60 keV with the sample cooled to cryogenic temperatures (T $\approx$ 150 K) as depicted in the schematic shown in Fig. \ref{monolayer_opt}(a). High-angle annular dark-field (HAADF) imaging encompasses Rutherford scattering towards high angular ranges whose cross-section approximates proportionally to the atomic-number (Z) by $\sim$Z$^{1.7}$. STEM-HAADF imaging of such freestanding WSe$_{2}$ monolayers, shown in the inset of Fig. \ref{monolayer_opt}(d), demonstrates the distinguishable intensity difference between the W and Se$_{2}$ atomic columns.

The EELS loss function, Im${\{-1/\epsilon\}}$, of layered materials like TMDCs  depends on the in-plane component of its complex dielectric function $\epsilon = \epsilon_{1} + i\epsilon_{2}$ \cite{Tizei2015}. As thickness approaches atomically-thin towards monolayers, the loss function becomes dominated by surface effects, and therefore it approximates to Im${\{\epsilon\}}$. It is thus equivalent to measuring the 2D material's absorption function ($\alpha$), and can reflect the optical excitation response, including the optical bandgap, and interband transitions. The excitonic absorption signatures of both freestanding and \textit{h}-BN encapsulated WSe$_{2}$ monolayers from low-loss EELS with the elastic (zero-loss) peak background subtracted demonstrates a good general correspondence to the optical absorption spectrum in Fig. \ref{monolayer_opt}(d)\cite{Arora2015,Schmidt2016}. With the exception of the additional broadening in the freestanding WSe$_{2}$ \cite{Shao2022}, the four dominant peaks, labeled as A, B, C, and D, are all well-reproduced in shape and energy positions. The theoretical understanding of the physical origin of each of these resonances in WSe$_{2}$ monolayer, including the high-energy spectral features, will be discussed in Section \ref{theory}.

\begin{figure}[tb]
	\centering
	\includegraphics[width=\columnwidth]{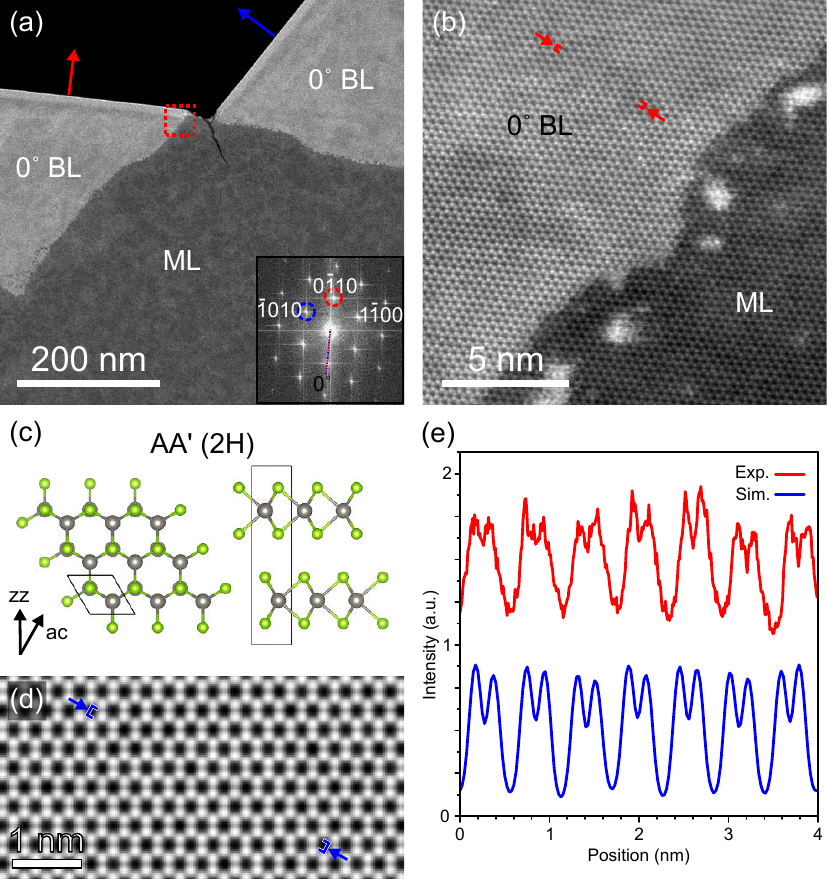}
	\caption{(a,b) STEM-HAADF images of WSe$_{2}$ monolayer (ML) folded along a zig-zag direction direction indicated in the inset image fast-Fourier transform in (a), resulting in 0$^{\circ}$ bilayers with AA$^{\prime}$ (2\textit{H}) stacking order. (c) Atomic model for AA$^{\prime}$ stacked WSe$_{2}$ bilayer with armchair (ac) and zig-zag (zz) directions noted in the projected view, and (d) multislice STEM-HAADF image simulation corresponding to the atomic model. (e) Intensity line profiles comparing experimental and simulated images along the selected areas marked by red and blue square brackets in (b) and (d), respectively. \label{AA-prime_stacking}}
\end{figure}

\begin{figure*}[tb]
	\centering
	\includegraphics[width=0.9\textwidth]{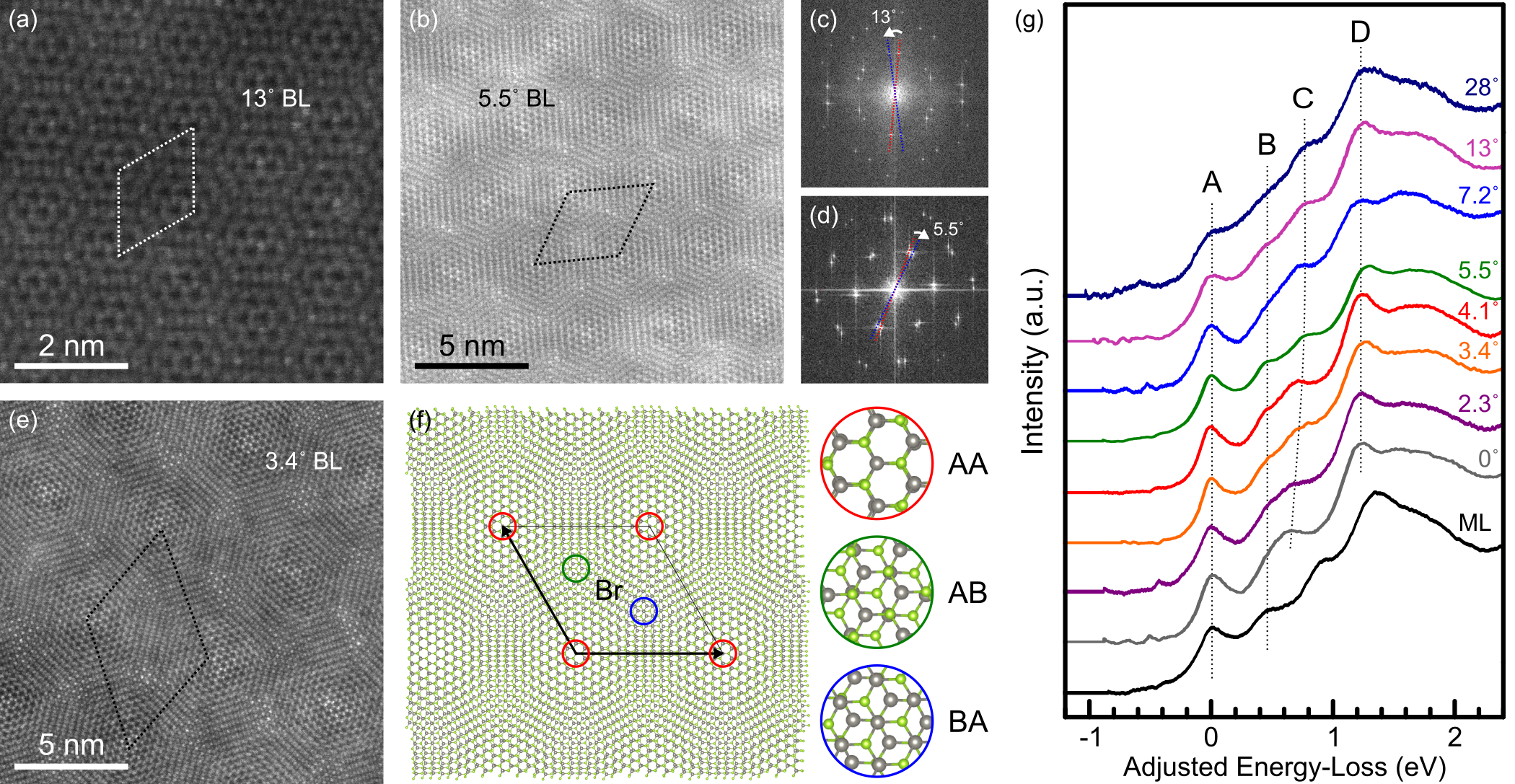}
	\caption{STEM-HAADF images of WSe$_{2}$ bilayers with (a) 13$^{\circ}$, (b) 5.5$^{\circ}$, and (e) 3.4$^{\circ}$ relative twist angle as measured from the image FFT of the larger twist angles in (c), (d), respectively. The moir\'{e} unit cells are highlighted by the dotted lines in (a), (b), and (e). In the image FFT, the red marks the orientation of the underlying monolayer, and the blue marks the orientation of the additional layer. (f) Atomic structure model of a moir\'{e} superlattice of twisted bilayer WSe$_{2}$ of 3.9$^{\circ}$ twist angle and \textit{R}-type stacking, with the moir\'{e} unit cell outlined in black, the various high-symmetry stacking identified by colored circles, and bridge site (Br) labelled. (g) Monochromated EELS spectra from twisted bilayer WSe$_{2}$ with various moir\'{e} angles compared to a representative monolayer (ML) offset in energy relative to the A exciton. The dotted lines are a guide to show the invariance and changes in the different exciton energy positions. \label{moire-spectra}}
\end{figure*}

In addition to freestanding WSe$_{2}$ monolayers, sub-micron fragments of bilayers (BLs) and trilayers (TLs) with variable relative twist angle between 0--30$^{\circ}$ were also common occurrences due to folding during the mechanical exfoliation and transfer process [\textit{cf.} Fig. \ref{monolayer_opt}(c) and Fig. \ref{folding-stack}]. The nature by which bilayers were formed can help shed light on their stacking order. In the case of those formed by folding of free edges, folding of monolayers along a zig-zag direction results in a bilayer with an aligned configuration of 0$^{\circ}$ relative twist angle and AA$^{\prime}$ stacking order (following the nomenclature proposed in ref. \onlinecite{He2014}), as shown in Fig. \ref{AA-prime_stacking}(a--c). AA$^{\prime}$ (2\textit{H}) bilayer stacking corresponds to the most energetically favorable configuration \cite{He2014}, typical of bilayers obtained by mechanical exfoliation from bulk crystals \cite{Sarkar2019}. STEM-HAADF image intensity was used to deduce the stacking in the bilayers by comparison with line profiles from multislice image simulations [Fig. \ref{AA-prime_stacking}(d,e)]. Fig. \ref{folding-stack}(e) illustrates schematically how folding of monolayers along a zig-zag direction (top), or in between zig-zag and armchair directions (bottom) can result in bilayers of zero and non-zero twist angles, respectively. The high twist angles of the 13$^{\circ}$ and 28$^{\circ}$ bilayers summarized in Fig. \ref{folding-stack} correspond to commensurate moir\'{e} angles with some of the smallest coincidence site lattices in homobilayers \cite{Zhao2022}, and falls under the second category when assessing the crystallography of the fold. The second-order reflection (armchair direction) circled in red in Fig. \ref{folding-stack}(d) indicates that for twist angle of 13$^{\circ}$, the folding normal (marked by the green arrow) lies closer towards an armchair direction, such that this would result in \textit{R}-type (or AA) stacking in the bilayer. Other notable features include the seldom Se-vacancy in monolayers \cite{Zheng2019}, and bands of oxide product made up of atomic clusters of tungsten from preferential oxidation at flake edges [\textit{cf}. Fig. \ref{AA-prime_stacking}(b), Fig. \ref{folding-stack}(c,g)].

The non-zero relative twist angles routinely observed vary from high twist angles [Fig. \ref{moire-spectra}(a) and Fig. \ref{folding-stack}] that generate sub-nanometer moir\'{e} periodicities to $\sim$10 nm periods for few degree twists [Fig. \ref{moire-spectra}(b,e)]. The relative twist angles were measured from the image fast Fourier-transform (FFT), including examples shown in Fig. \ref{moire-spectra}(c,d), and confirmed using nano-beam electron diffraction, where the latter is less sensitive to effects of scan distortion in STEM imaging. Well-defined hexagonal moir\'{e} patterns with few-nanometer periodicity are also evident in the STEM-HAADF images for the low twist angles in Fig. \ref{moire-spectra}(b,e) and Fig. \ref{3layer_low-angle}. The moir\'{e} superlattice (as highlighted by a dotted line in the images) of the WSe$_{2}$ homobilayers is composed of regions of high-symmetry stacking, namely, AA, AB, and BA stacking in the case of \textit{R}-type stacking \cite{Gogoi2019}. As shown in the moir\'{e} superlattice structure model in Fig. \ref{moire-spectra}(f), it is the bridge sites (marked Br) connecting adjacent AB and BA stacked regions, which also have their own unique local alignment, that make up the hexagonal pattern outlines of the twisted bilayers observable in the STEM-HAADF images. Unlike the common AA$^{\prime}$ stacking, these aforementioned \textit{R}-type stacking configurations lack inversion symmetry, such that the \textit{K} ($K^{\prime}$)-points of the joint Brillouin zone are inequivalent and the spins of the upper and lower split bands in individual layers are instead identical at a given valley\cite{Schneider2019}. For twisted bilayers, the energy of the indirect $KQ$ transition has been shown to depend on twist-angle and atomic registry in MoS$_{2}$ \cite{vanderZande2014}, WS$_{2}$ \cite{Zheng2015}, and more recently in WSe$_{2}$ \cite{Wang2018,Scuri2020,Merkl2020}.

Low-angle annular dark-field (LAADF) imaging in STEM offers more diffraction contrast sensitivity, and can be used to image the moir\'{e} superlattice over long-range (hundreds of nm), thus particularly useful to illustrate distortions in the local periodicity over long distances \cite{Weston2020}. At twist angles towards 5$^{\circ}$, STEM-LAADF imaging gives a periodic pseudo-atom like contrast that corresponds to the various high-symmetry stacking points in the moir\'{e} superlattice. Distortions in the moir\'{e} lattice at ripples, cracks or towards edges of the twisted WSe$_{2}$, such as shown in Fig. \ref{atomic-reconstruction}(g,i), lead to release of strain and thus reconstruction into domains (an expansion of specific high-symmetry points). The domain contrast is further manifested at the lowest twist angle of 2.3$^{\circ}$, where arrays of triangular domains with boundaries in dark contrast are arranged in a six-fold fashion [as marked by alternating purple and turquoise triangles in Fig. \ref{atomic-reconstruction}(c)] continuous over the entirety of the few hundred nm-sized twisted bilayer and trilayer. The domain boundary geometry differs between \textit{R}-type (AA) and \textit{H}-type (AA$^{\prime}$) stacking in twisted homobilayer TMDCs \cite{Weston2020}, taking on a triangular geometry and kagome-like pattern dominated by hexagonal regions, respectively. Both geometries are governed by atomic reconstruction where some of the high-symmetry stacking regions become more energetically favorable towards low twist angles.

The purpose of the STEM-LAADF imaging on twisted WSe$_{2}$ bilayers is effectively two-fold: firstly to identify lattice distortions and the occurrence of atomic reconstruction; and secondly, the domain boundary geometry was used to identify the stacking order in lower twist angles. Three of the five twisted bilayers formed by stacking presented in Fig. \ref{atomic-reconstruction} have been identified to have \textit{R}-type (AA) stacking. In addition, comparative STEM-HAADF images of regions with atomic reconstruction [see Fig. \ref{atomic-reconstruction}(d) and (h)] confirms the domain boundary are the bridge sites (Br), and the triangular domains are made up of AB/BA stacking configuration as outlined by the purple triangle in Fig. \ref{atomic-reconstruction}(d). Observation of atomic reconstruction over long-range in only the 2.3$^{\circ}$ twisted bilayer WSe$_{2}$ validates the calculated crossover angle for \textit{R}-type stacked bilayer TMDCs of $\theta^{\circ}_{3R}$ $\sim$ 2.5$^{\circ}$ by Enaldiev \textit{et al}.\cite{Enaldiev2020}, below which the bilayers transition from a rigid rotation to a lattice-reconstructed regime.

The excitonic absorption signatures of twisted WSe$_{2}$ bilayers from low-loss EELS are presented in Fig. \ref{moire-spectra}(g) aligned in energy with respect to the A exciton alongside a spectrum from a WSe$_{2}$ monolayer for comparison. The four excitonic resonances are also prominently reproduced in the case of twisted bilayers. The energy separation between A and B exciton listed in Table \ref{exciton-energies}, which is linked to the spin-orbit coupling, remains relatively constant with the number of layers, as well as twist angle. Small shifts in the A exciton, coupled with simultaneous rigid shifts of the B exciton are visible in the unadjusted spectra in Fig. \ref{images_spectra}(c) and they can be attributed to local strain \cite{Schmidt2016} or unintentional doping. It is worth noting the small peaks $\sim$300 meV below the A exciton [Fig. \ref{moire-spectra}(g) or at 1.3--1.4 eV as presented in Fig. \ref{images_spectra}(c)] is likely of the same origin as the so-called subgap exciton peak measured using momentum (q)-resolved EELS at non-zero q in various TMDCs including WSe$_{2}$ \cite{Hong2020}.

\begin{table*}
  \sisetup{round-mode=places,round-precision=3}
  \centering
  \caption[]{Fitting of the A, B, C, D excitonic peak energy positions (X$_{A}$, X$_{B}$, X$_{C}$, X$_{D}$) obtained from the EELS spectra in Fig. \ref{images_spectra}(c), and the relative energy difference between A and B exciton (A--B) governed by valence band splitting at $K$-point, B and C exciton (B--C), and A and C exciton (A--C). Twist angles marked by asterisks (*) are relative twist angles because its stacking order remains undetermined and can also be 60 -- $\theta$.}
  \label{exciton-energies}
  \small
  \begin{ruledtabular}
 \begin{tabular}{cccccccc}
    Twist angle ($^{\circ}$) & X$_{A}$ (eV) & X$_{B}$ (eV) & X$_{C}$ (eV) & X$_{D}$ (eV) & $\Delta_{A-B}$ (eV) & $\Delta_{B-C}$ (eV) & $\Delta_{A-C}$ (eV) \\
    \hline
	Monolayer & \num{1.69665} & \num{2.14549} & \num{2.59033} & \num{3.02347} & 0.449 & 0.445 & 0.894\\
	0 & \num{1.75122} & \num{2.24806} & \num{2.38763} & \num{2.95003} & 0.497 & 0.140 & 0.636\\
	2.3 & \num{1.75306} & \num{2.22515} & \num{2.41646} & \num{2.95970} & 0.472 & 0.191 & 0.663\\
	3.4 & \num{1.71144} & \num{2.16072} & \num{2.42001} & \num{2.93420} & 0.449	 & 0.259 & 0.709\\
	4.1* & \num{1.74908} & \num{2.19568} & \num{2.44703} & \num{2.97394} & 0.447 & 0.251 & 0.698\\
	5.5 & \num{1.71309} & \num{2.15818} & \num{2.48009} & \num{2.97057} & 0.445 & 0.322 & 0.767\\
	7.2* & \num{1.76281} & \num{2.22385} & \num{2.48613} & \num{2.95203} & 0.461 & 0.262 & 0.723\\
	13 & \num{1.69722} & \num{2.14272} & \num{2.47160} & \num{2.93910} & 0.446 & 0.329 & 0.774\\
	28 & \num{1.74317} & \num{2.18494} & \num{2.51673} & \num{2.98775} & 0.442 & 0.332 & 0.774\\
  \end{tabular}
  \end{ruledtabular}
\end{table*}

Comparing different twist angles in the bilayers also show sizable shifts in the C exciton energy up to 200 meV, which subsequently alter drastically the overall shape of the spectrum at the B--C transitions, with extremes between the aligned (0$^{\circ}$ and 60$^{\circ}$) and anti-aligned (towards 30$^{\circ}$) cases suggesting underlying differences in interlayer coupling. The exciton peak shifts are quantitatively determined by peak fitting to the second-derivative treated with Savitzky-Golay filtering using multiple Gaussians, four in total, each corresponding to a structure in each EELS spectrum. The results of the peak fitting are summarized in Table \ref{exciton-energies} and displayed graphically in Fig. \ref{images_spectra}(e). Consistent with optical absorption \cite{Arora2015,Zhao2013}, the layer thickness dependence of A, B, and C exciton resonances of few-layered WSe$_{2}$ measured using monochromated EELS shows a pronounced decrease in the C and D exciton energies between the monolayer and 0$^{\circ}$ bilayer with AA$^{\prime}$ stacking in Fig. \ref{moire-spectra}(c). 

Furthermore, the C exciton energy continues to show the same decreasing trend with the number of layers when comparing bilayers and trilayers of the same twist angle [Fig. \ref{3layer_low-angle}(b)]. The more pronounced shifts of the C and D excitonic peaks relative to the A and B excitons with layer thickness suggests an association to the localization of the electronic states involved in the respective transitions, in particular the orbital character of the chalcogen atoms (Se $p$-orbitals in this case) that contribute most to the interlayer coupling \cite{Zhang2015orbitals}. Specifically, the valence band maximum (VBM) at $K$-point exhibits in-plane $p_{x,y}$ character, while the VBM and conduction band minimum (CBM) towards $\Gamma$-point displays mainly out-of-plane $p_{z}$ character \cite{Voss1999} and thus most strongly affected by interlayer separation in few-layered WSe$_{2}$; the $p$-orbital contribution shows a mixture of $p_{x,y}$ and $p_{y}$ character at $Q$-point [see the orbital-projected band structure for AA$^{\prime}$ WSe$_2$ in Fig. \ref{orbitals}].
This corroborates well with the expected changes in the band structure between monolayer towards bulk WSe$_{2}$, namely the appearance of an indirect gap $K$-$Q$ transition due to the downshift of the $Q$-valley overtaking the $K$-point as CBM beyond a monolayer. For the twisted bilayer WSe$_{2}$ with increasing relative twist angle, the C exciton energy blueshifts by $\sim$200 meV [\textit{cf.} Table \ref{exciton-energies} and Fig. \ref{images_spectra}(e)], indicative of an upshift in the CBM at $Q$-valley. Recent studies on twisted bilayer WSe$_{2}$ reported similar blueshifting in the indirect $KQ$ interlayer exciton emission energy as a function of twist angle, reaching a maximum at $\theta$ = 30$^{\circ}$, while relatively minimal change in the direct $KK$ intralayer exciton (X$_{A}$) in comparison \cite{Scuri2020,Merkl2020}. Therefore the phonon-assisted indirect $KQ$ exciton energy directly reflects the interlayer electronic coupling strength, which is strongest at 0$^{\circ}$ and 60$^{\circ}$ \cite{Liu2014}. Raman spectroscopy is typically used as an indicator of the mechanical (i.e. vibrational) interlayer coupling \cite{Castellanos-Gomez2014,Liu2014,Sarkar2019}.
In a similar manner, energy shifts of the C exciton energy can also gauge the electronic interlayer coupling effects from absorption-based techniques such as EELS, indicating a reduction in its strength with moir\'{e} angle towards 30$^{\circ}$ in twisted WSe$_{2}$ bilayers.

\section{Discussion}
\label{theory}

Fundamental insights on the excitonic response of TMDCs can be successfully acquired using the \textit{GW}+BSE (Bethe-Salpeter equation) method on top of DFT calculations \cite{Komsa2012,Qiu2016,Marsili2021}. In the case of W-based TMDC monolayers, the lowest energy exciton A is mainly composed of transitions near the $K$ point from the VBM to the second unoccupied state in the conduction band (CBM+1) which has the same spin-character \cite{Hong2021,Gillen2021,Marsili2021}. The B peak has a more complex character but it is mostly formed by transitions near $K$ from VBM$-$1 to CBM. As discussed in the literature \cite{Hong2021,Gillen2021}, higher energy spectral features cannot be linked to individual excitons but arise from an ensemble of excitonic transitions very close in energy. The complex TMDCs excitonic spectra is therefore usually broadened to reproduce the same number of peaks seen in experiments. However these structures may not present a homogeneous excitonic character and result from a superposition of excitons belonging to different orders of distinct Rydberg series. This situation has led to an uneven nomenclature \cite{Zhao2013,Kozawa2014,Gillen2017IEEE,Hong2021,Gillen2021}, and particular attention should be paid when comparing different references.

\begin{figure}
	\includegraphics[width=\columnwidth]{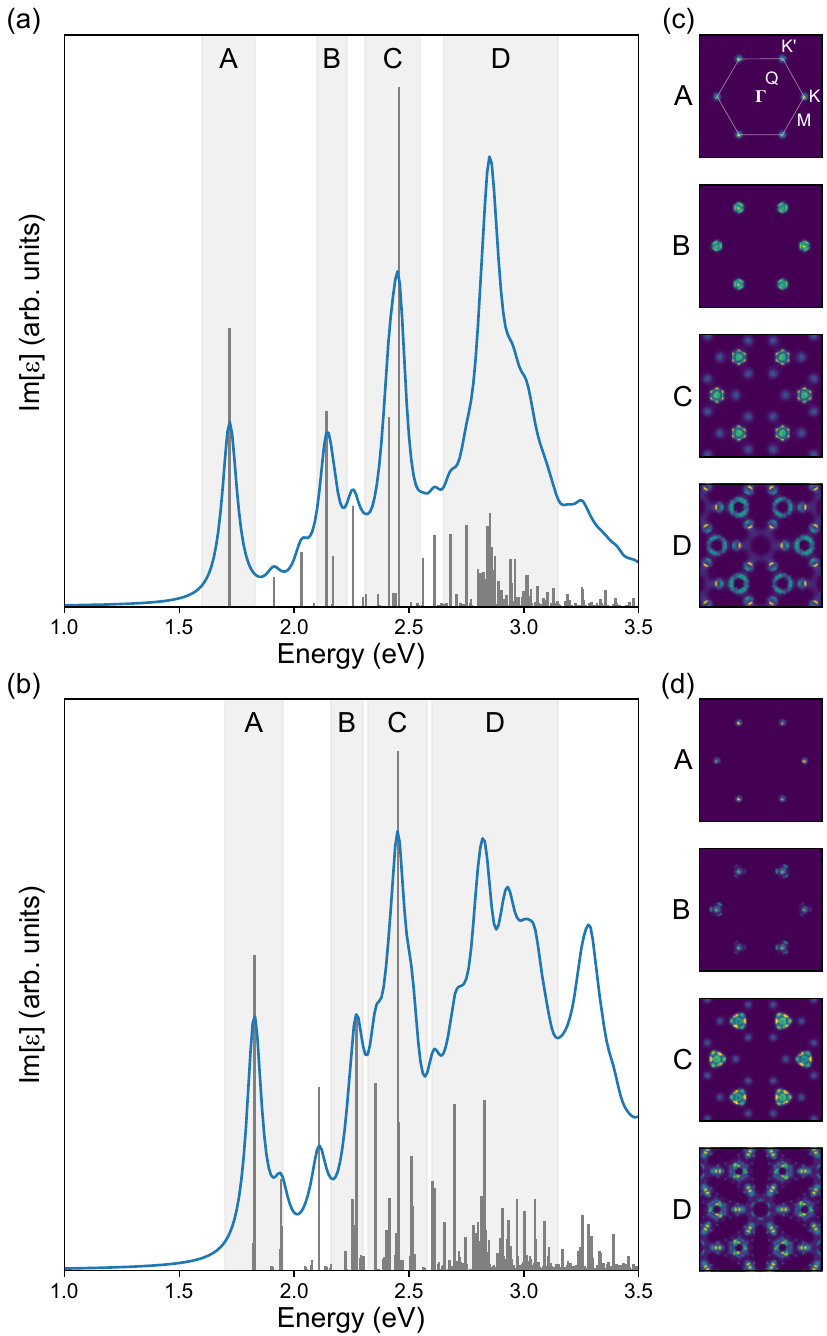}
	\caption{Imaginary part of the dielectric function calculated using a GW+BSE approach for (a) monolayer and (b) AA$^\prime$ bilayer \ce{WSe2} together with the oscillator strengths of the main excitonic transitions. The shaded regions correspond to the main peaks identified in experiments. (c,d) Weight in reciprocal space of the transitions contributing to each of these peaks.}%
	\label{spectra}
\end{figure}

The character of the excitonic transitions in the AA$^{\prime}$ bilayer cannot be deduced \textit{a priori} from the spectroscopic response of the monolayer. Therefore, the AA$^{\prime}$ bilayer is treated explicitly, along with the monolayer case, via a non-collinear \textit{GW}+BSE approach (computational details can be found in the SI). This method has been shown to provide good agreement with experiments for the energy separation of the A and B excitons compared to a perturbative treatment of spin-orbit coupling \cite{Marsili2021}. The imaginary part of the dielectric function $\epsilon_2$ for the monolayer and AA$^\prime$ bilayer \ce{WSe2} together with the oscillator strengths of the main excitonic transitions are shown in Fig. \ref{spectra}(a) and (c). For the sake of comparison with experiments, four energy windows centered at local maxima of $\epsilon_2$ have been defined for the mono- and bilayer spectra, respectively, which can be linked to the experimental peaks A--D [shaded regions in Fig. \ref{spectra}(a,b)]. Fig. \ref{spectra}(c,d) presents the weight in the reciprocal space of the transitions contributing to each of these peaks: for each exciton $\lambda$ in a given energy window, the weights $\sum_{vc}A^{cvk}_\lambda$ of the electron-hole pairs of wave vector $k$ are considered, and all these contributions are summed up taking into account the oscillator strength of each individual exciton. 

Peaks A and B are formed by transitions near $K$ and both features undergo a blueshift in the bilayer due to an increase in the direct band gap at $K$ compared to that of the monolayer from. Peak C has a more complex structure where \textit{k}-points next to both $K$ and $Q$ contributes. While the points next to $K$ have a higher spectral weight, the $Q$ points are three times more numerous and thus the integrated contribution of the two regions of the reciprocal space is comparable. Peak C had been previously described as a higher-order exciton of the same series as peak B \cite{Hong2021}; Fig. \ref{spectra}(c,d) illustrates that additional excitonic transitions contribute to this spectral feature. Finally, peak D is dominated by transitions from the last occupied to the first unoccupied band near the $Q$ point. The appearance of these strong high-energy excitonic transitions had been linked to the high joint density of states that arises from band nesting effects in TMDCs \cite{Carvalho2013,Bieniek2018,Mennel2020}.

As discussed in the experimental results, blueshifts of the excitonic peak C in bilayer TMDCs were observed as a function of their twist angle by means of EELS. It can be reasonably argued that the decomposition in the reciprocal space of these spectral features might be the same for aligned and twisted bilayers. Therefore, while the \textit{GW}+BSE approach gets too computationally expensive when applied to extended moir\'{e} supercells, it might be feasible to link trends observed in the spectra to continuous changes of the DFT electronic structure. The band structure of a moir\'{e} supercell is highly folded and therefore it can be hardly compared to those of a reference untwisted bilayer or monolayer without the use of unfolding methods which provide a primitive cell effective band structure \cite{Ku2010,Lee2013unfolding}. These techniques have been employed already for the unfolding of the bands of various twisted 2D heterostructures \cite{Nishi2017,Matsushita2017,Sanchez-Ochoa2020,Magorrian2022band}. Unfolding routines require the definition of a reference primitive basis; in the case of moir\'{e} structures, unfolding has to be performed twice to take into account the intrinsic periodicity of each of the two layers separately. The unfolded bands can then be projected independently on the two layers used as reference.

\begin{figure*}[!htb]
	\includegraphics[width=0.95\textwidth]{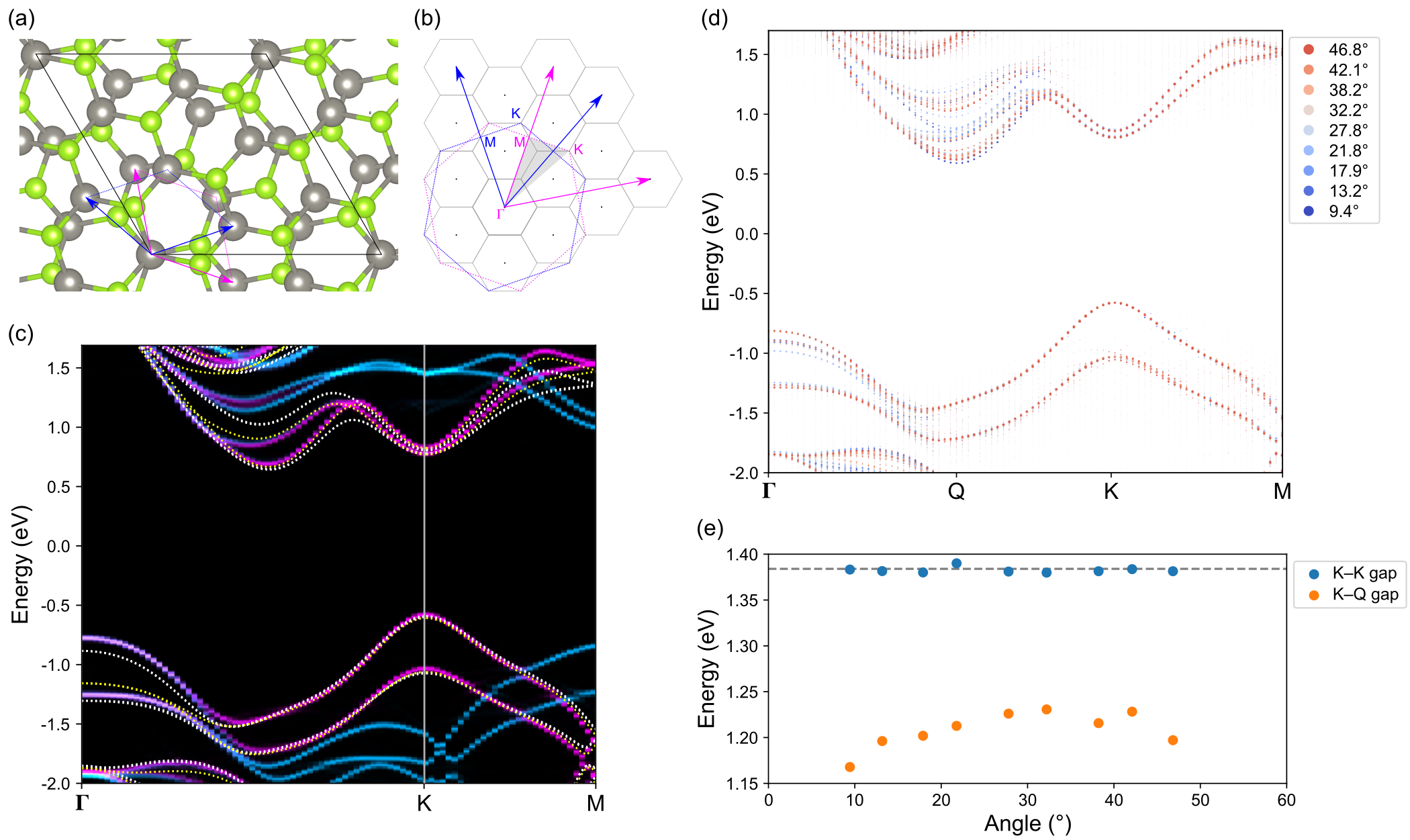}
	\caption{(a) Structure of a WSe$_{2}$ moir\'{e} supercell with a twist angle of 21.8$^\circ$. Blue and purple arrows indicate the primitive unit cell of the top and bottom layer, respectively. (b) Reciprocal lattice vectors and Brillouin zones of the primitive cells of the individual layers, and the supercell Brillouin zones (grey hexagons). (c) DFT band structure of the moir\'{e} cell unfolded along the high symmetry directions of the irreducible Brillouin zone shaded in (b). Blue and purple lines correspond to the unfolding using as reference primitive cell those of the top and bottom layer, respectively. The yellow and white dashed lines are the band structure of WSe$_{2}$ monolayer and untwisted bilayer, respectively. (d) Unfolded band structures of WSe$_{2}$ moir\'{e} supercells with different twist angles projected over the reference layer. (e) Direct and indirect band gap as a function of the twist angle. Gray dashed line indicates the direct gap of the WSe$_{2}$ monolayer.}
	\label{unfolded-example}
\end{figure*}

As an example, in Fig. \ref{unfolded-example}(a) the unfolding method has been applied to a moir\'{e} supercell with a twist angle of 21.8$^\circ$ (computational details are provided in the SI). The purple and blue bands in Fig. \ref{unfolded-example}(c) were unfolded using the primitive cell of the bottom and top layer as a reference [following their color-coded cells in Fig. \ref{unfolded-example}(b)], respectively, and were subsequently projected onto the same layer. The $\Gamma$--$K$--$M$ path connects high symmetry points of the bottom layer but not of the top layer. The unfolded occupied purple bands closely follow those of the untwisted AA$^\prime$ bilayer (white dashed lines), although an upper shift is observed near the $\Gamma$ point. This region of reciprocal space is where the occupied bands of the monolayer (yellow dashed line) and bilayer differ and is very sensitive to the interlayer spacing which, in twisted TMDCs, can vary with the twist angle and the layer registry \cite{vanderZande2014,Yan2019}. Finally, a few small minigaps ($< 50$ meV) open at the crossing of the bands of the two layers if they present the same spin character. More relevant variations are observed in the conduction bands. With respect to the AA$^\prime$ bilayer, the first two unoccupied bands of the monolayer cross in the $K$--$Q$ path and are higher in energy at $M$. The unfolded purple bands show the same characteristics and can therefore be reasonably interpreted as the bands of the monolayer perturbed by the adjacent twisted layer.

In Fig. \ref{unfolded-example}(d) the unfolded bands of twisted \ce{WSe2} bilayers are presented as a function of the twist angle. At the $K$ point, both the direct band gap and the spin-orbit splitting do not change with the twist angle and remain equal to the value of the untwisted bilayer [blue dots in Fig. \ref{unfolded-example}(e)]. This behavior can be explained by the reduced interlayer orbital coupling for the band-edge states at the $K$ point \cite{Kang2016unified}. Reasonably assuming that layer twists will only have a minor effect on screening, invariance of low-energy excitonic features can be deduced, as observed for the A and B peaks in the EELS spectra in Fig. \ref{moire-spectra}(g).

While the valence band along the $K$--$Q$ path remains invariant with the twist angle, the bottom of the CBM near $Q$ point upshifts. The values of the indirect gap $KQ$ as a function of the twist angle were extracted and plotted in Fig. \ref{unfolded-example}(e). The maximum of the indirect gap occurs for twist angles close to 30$^\circ$ and progressively decreasing of several tens of meV going towards 0$^\circ$ and 60$^\circ$. This behavior reproduces well the trend observed in the indirect gap measured by photoluminescence \cite{Merkl2020} where $KQ$ indirect excitons can be activated by phonons. C and D excitons observed in EELS, optical absorption or reflectivity, involves dipolar transitions near the $Q$-point. Since the upper valence band in this reciprocal space region is not affected by the twist, the energy difference of the vertical transitions follows the same trend as the indirect gap, but it is shifted at higher energies. While the excitonic response of twisted \ce{WSe2} has not here been explicitly calculated, the analysis of the unfolded bands combined with the study of the excitonic character of spectral features from mono and bilayer permits explaining the experimental trends observed in EELS as a function of twist angle.
It should be noted that this interpretative scheme may not be valid when considering very low (near-zero) twist angles, where the moir\'{e} structure undergoes extensive structural relaxations and where excitonic states may re-hybridise \cite{Brem2020}, giving rise to complex spectral features that cannot be simply linked to the spectral response of the perfect mono- or bilayer.

\section{Conclusions}
The evolution of the excitonic response in twisted bilayer WSe$_{2}$ as a function of moir\'{e} angle has been investigated using monochromated STEM-EELS under cryogenic conditions, highlighting a progressive blueshift of the high-energy C excitonic peaks relative to the AA$^{\prime}$-stacked bilayer. Atomically-resolved imaging was used to provide relevant structural information on the twisted bilayers, including the twist angle and stacking order, in addition to revealing the occurrence of atomic reconstruction in the lowest observed twist angle of 2.3$^{\circ}$. In combination with first-principles calculations based on the \textit{GW}+BSE approaches, the physical origin of the high-energy spectral features in monolayer and AA$^{\prime}$ bilayer WSe$_{2}$ were examined. Moreover, the unfolded DFT electronic structure of twisted bilayers showed an uplifting of the $Q$-valley CBM with respect to the untwisted AA$^{\prime}$ bilayer. The trends in band structure changes with moir\'{e} angle were then linked to the BSE calculated dielectric response of the untwisted bilayer, giving good agreement to the dipolar transitions near $Q$-point contributing to the high-energy C exciton observed in EELS from the current work as well as the phonon-assisted indirect $K$--$Q$ transition measured by photoluminescence by other groups. Therefore tuning of the C exciton transitions as measured by absorption spectroscopy like EELS is an effective indicator of the electronic interlayer coupling.

With capabilities to collect photons generated by cathodoluminescence within the electron microscope utilised in this study, the addition of \textit{h}-BN encapsulation can bring further insight to interlayer interactions in such twisted bilayers, in particular at the lengthscales of the moir\'{e} periodicity. The expected reduction in EELS absorption linewidths will aid in the identification of small spectral variations; in conjunction, sufficient excitation of electron-hole pairs in the \textit{h}-BN for recombination in the TMDC opens the possibility for concurrent correlation to the indirect exciton emission in the twisted bilayers.

\acknowledgments{
The authors acknowledge funding from the ANR, program of future investment TEMPOS-CHROMATEM (No. ANR-10-EQPX-50). This work was supported by the European Union in the Horizon 2020 Framework Program (H2020-EU) under Grant Agreement No. 823717 (ESTEEM3) and 101017720 (eBEAM). S.Y.W. acknowledges NSERC for the postdoctoral fellowship funding. A.A. acknowledges financial support from the German Research Foundation (DFG Project Nos. AR 1128/1-1 and AR 1128/1-2) and and NM-ICPS of the DST, Government of India through the I-HUB Quantum Technology Foundation (Pune, India). M.P. acknowledges CINECA for CPU time granted within ISCRA-B and ISCRA-C initiatives. This research was sponsored [in part] by the NATO Science for Peace and Security Programme under grant G5936.
}

\bibstyle{apsrev4-1}
%

\newpage
	\renewcommand{\thefigure}{S\arabic{figure}}
	\renewcommand{\thesubsection}{S\arabic{subsection}}
	\setcounter{figure}{0} 
	\setcounter{section}{0}	

\section{Supporting Information}

\subsection{STEM Characterization}
The atomically-resolved imaging has been performed on an aberration-corrected Nion UltraSTEM200 operated at 60 keV using a convergence angle of 33 mrad with typical beam currents of 20 -- 40 pA and a detector angular range of 80 -- 200 mrad for the high-angle annular dark-field (HAADF) images. Additional low-angle annular dark-field (LAADF), which mimicks the diffraction contrast of dark-field TEM for twisted bilayer graphene domains \cite{Brown2012_DF-TEM,Yoo2019_tBLgr} was carried out using 10 mrad convergence angle and 14 -- 28 mrad detector angular range. The monochromated EELS and electron diffraction were performed on a modified Nion HERMES-S200 (also known as ChromaTEM) operated at 60 keV with the sample at liquid nitrogen temperatures (T $\approx$ 150 K) using 10 -- 15 mrad and 3 -- 5 mrad convergence angle, respectively.
STEM-HAADF image simulations were performed using a multislice simulation package called QSTEM,\cite{Koch_QSTEM} with the input parameters (convergence angle, HAADF detector collection angles) set according to the experimental conditions.
The EELS spectra were processed using a combination of Gatan DigitalMicrograph, and the python package HyperSpy.\cite{delaPena_HyperSpy} Energy positions of exciton peaks were extracted by Gaussian peak fitting the negative of the second-derivative of smoothed spectra obtained by Savitzky-Golay filtering. All spectra displayed are in their raw state, summed over tens of nanometer regions with a power-law background subtracted.

\subsection{Computational Details}
\paragraph{Spectra simulations}

The mono- and bi-layer \ce{WSe2} structures have been optimized using the Quantum Espresso code\cite{giannozzi2009quantum} using the Perdew-Burke-Ernzerhof (PBE) exchange-correlation functional\cite{PBE} and optimized norm-conserving pseudopotentials.\cite{Hamann-13} Dispersion forces have been introduced to relax both the mono and bilayer by the vdW revised-DF2 method which has been demonstrated to be accurate for 2D materials.\cite{Hamada-14}
DFT eigenvalues and eigenfunctions have been used in successive many-body simulations performed using the YAMBO code.\cite{yambo09,yambo19} Quasi-particle energies have been calculated via a GW perturbative one-shot method and optical excitation energies and optical spectra by solving the Bethe-Salpeter Equation. Further details on the computational parameters employed both in DFT and many-body calculations can be found in Ref. \citenum{Marsili-21}, appendix D. The analysis of the weight in the reciprocal space of the optical transitions contributing to the excitons has been later performed using the yambopy post-processing package.\cite{yambo19}

\paragraph{Band unfolding} DFT calculations in the band unfolding part have been performed using the OpenMX code.\cite{Ozaki2003} The exchange-correlation potential was expressed in the generalized gradient approximation using the Perdew-Burke-Ernzerhof (PBE) schema. Norm conserving full relativistic pseudopotentials including a partial core correction and spin-orbit coupling was used. Dispersion interactions have been included using the Grimme-D2 method.\cite{grimme2010} A basis set of optimized numerical pseudo-atomic orbitals was employed as provided by the openmx 2019 database.

\bibstyle{apsrev4-1}
%

\begin{figure*}
\includegraphics[width=0.95\textwidth]{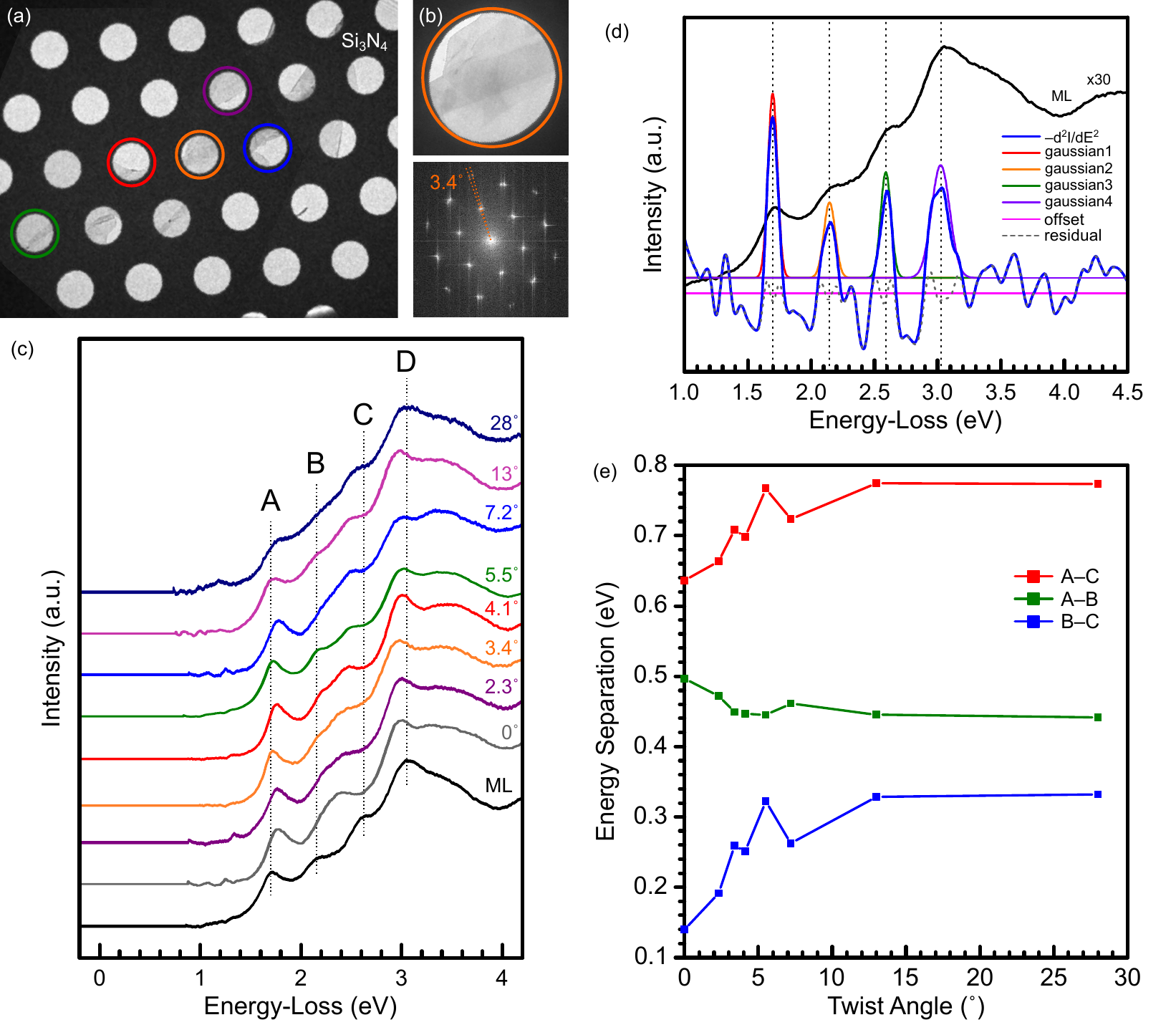}
\caption{\label{images_spectra} (a) Low-magnification STEM bright-field (BF) image of WSe$_{2}$ layers with many holes marked containing few hundred nanometer-sized fragments of bilayers (BLs) and trilayers (TLs) with variable twist angles, including that in (b) with 3.4$^{\circ}$ twist angle measured from the image fast-Fourier transform. (c) Monochromated EELS spectra from twisted bilayers with various moir\'{e} angles compared to a representative monolayer (ML). (d) Example of the fitting procedure of the exciton peak positions by Gaussian peak fitting of the $-d^{2}I/dE^{2}$ on Savitzky-Golay filtered spectra. (e) Graphical representation of the transition energy difference between select exciton resonances as a function of twist angle from the peak fitting.}
\end{figure*}

\begin{figure*}
\includegraphics[width=\textwidth]{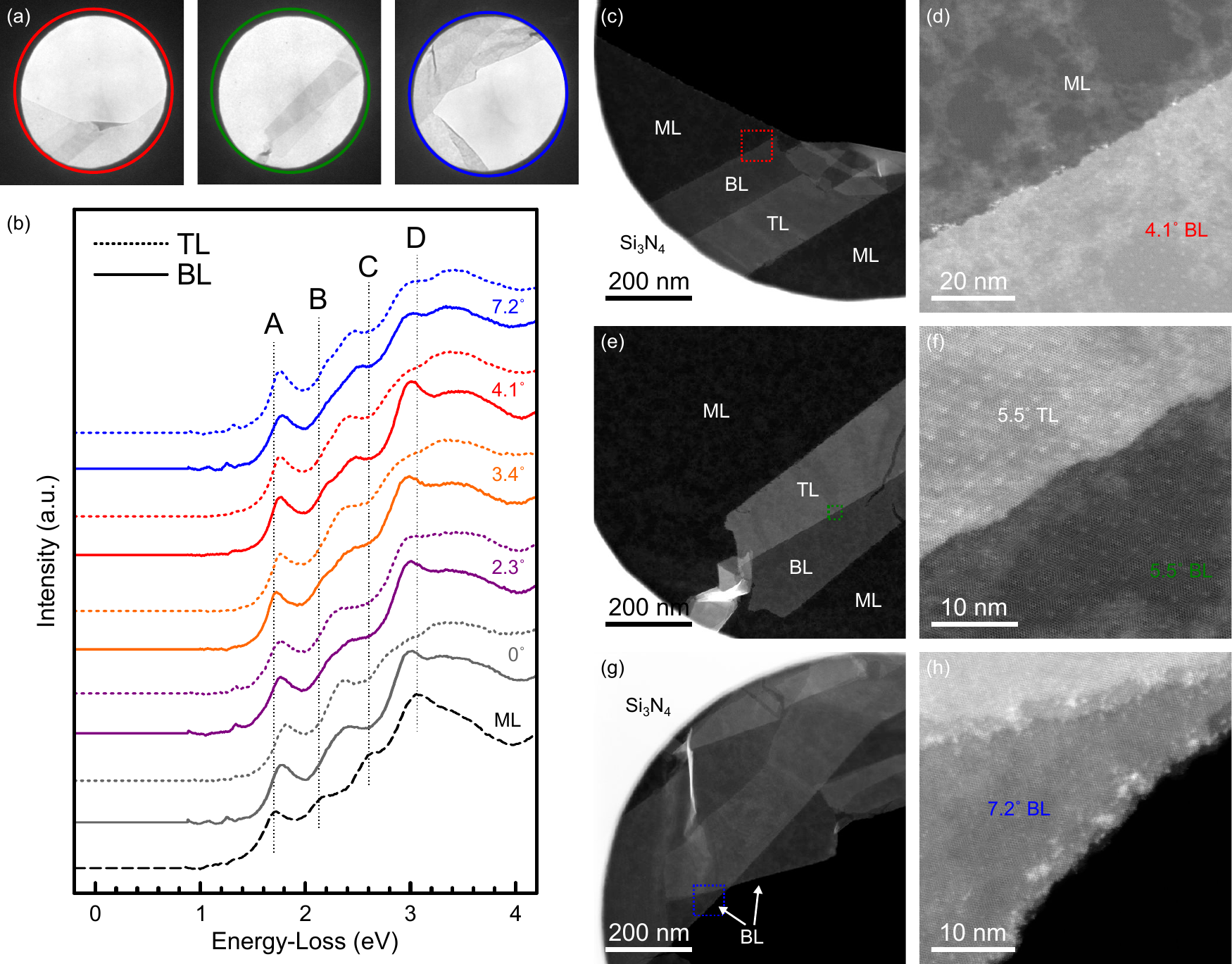}
\caption{\label{3layer_low-angle}(a) STEM bright-field (BF) images of WSe$_{2}$ layers from holes marked in Fig. \ref{images_spectra} following the same color-code. (b) Monochromated EELS spectra from twisted BLs and TLs with various moir\'{e} angles, including 0$^{\circ}$, compared to a representative ML. STEM-HAADF images of the WSe$_{2}$ layers from the holes marked in red (c,d), green (e,f) and blue (g,h) from the BF images in (a), highlighting the different moir\'{e} periodicity with twist angle in the high-magnification images in (d,f,h).}
\end{figure*}

\begin{figure*}
\includegraphics[width=\textwidth]{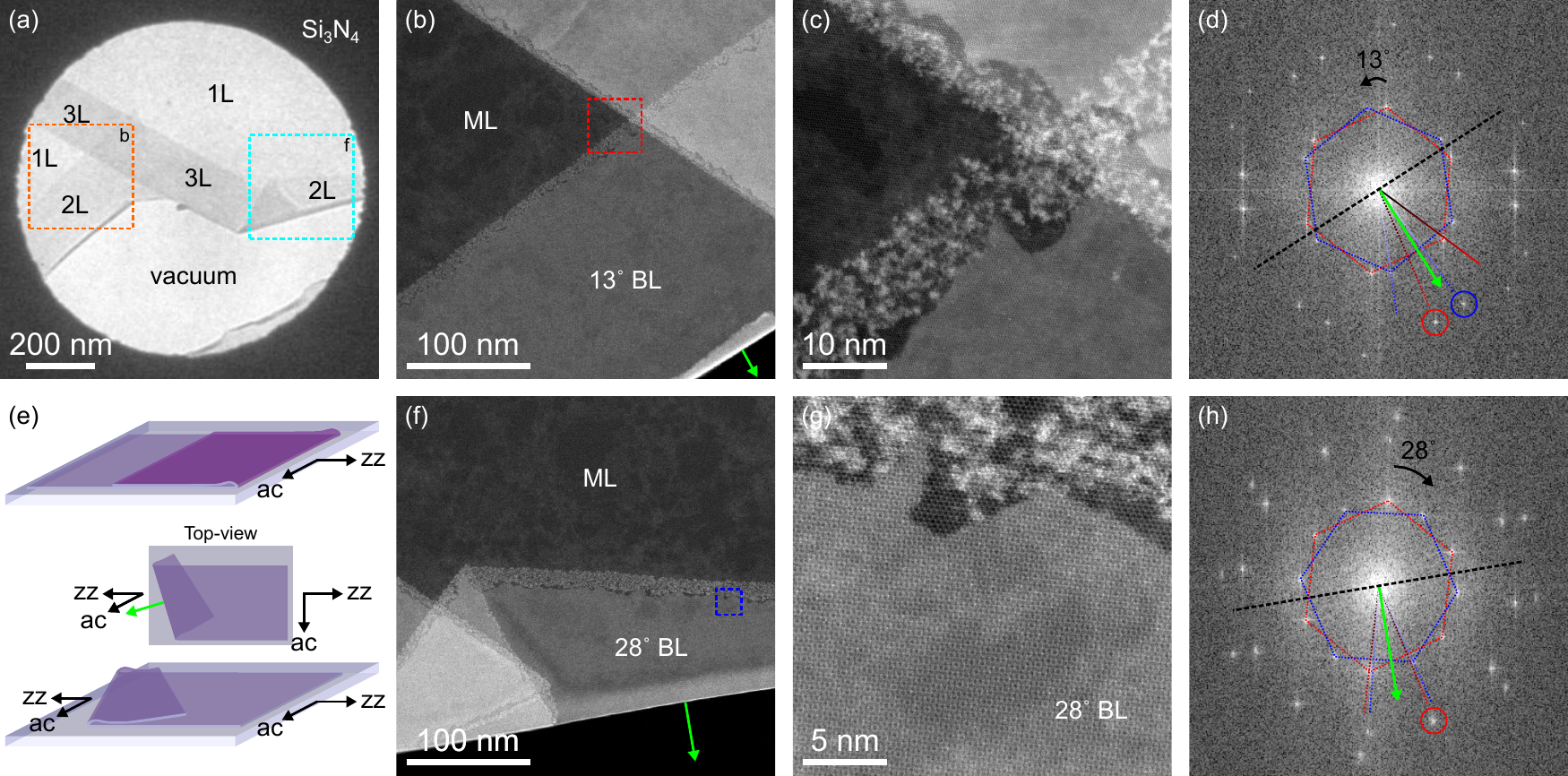}
\caption{\label{folding-stack}(a) Low-magnification STEM bright-field (BF) image of WSe$_{2}$ layers including the two BL regions with high twist-angle formed by folding from Fig. \ref{moire-spectra}(a). STEM-HAADF images of the WSe$_{2}$ layers with (b--c) 13$^{\circ}$ and (f--g) 28$^{\circ}$ twist angle from the regions marked correspondingly in the BF image in (a). The image FFT with the assessment of the folding orientation (black dashed line) and its normal direction (green arrow) with respect to the closest armchair and zig-zag directions (dotted lines) of the underlying ML for (d) 13$^{\circ}$, and (h) 28$^{\circ}$ twist angles. The red-dashed hexagon marks the orientation of the underlying ML, and the blue marks the orientation of the additional layer. The second-order reflection (armchair) circled in red indicates that for both angles, the folding normal lies towards an armchair direction, such that this would result in AA (or R-type) stacking in the BL. (e) Schematic illustration of how folding of MLs along a zig-zag (zz) direction [top] such as in Fig. \ref{AA-prime_stacking}, or in between zig-zag (zz) and ac directions [bottom] can result in BLs of zero or non-zero twist angles, respectively.}
\end{figure*}

\begin{figure*}
\includegraphics[width=\textwidth]{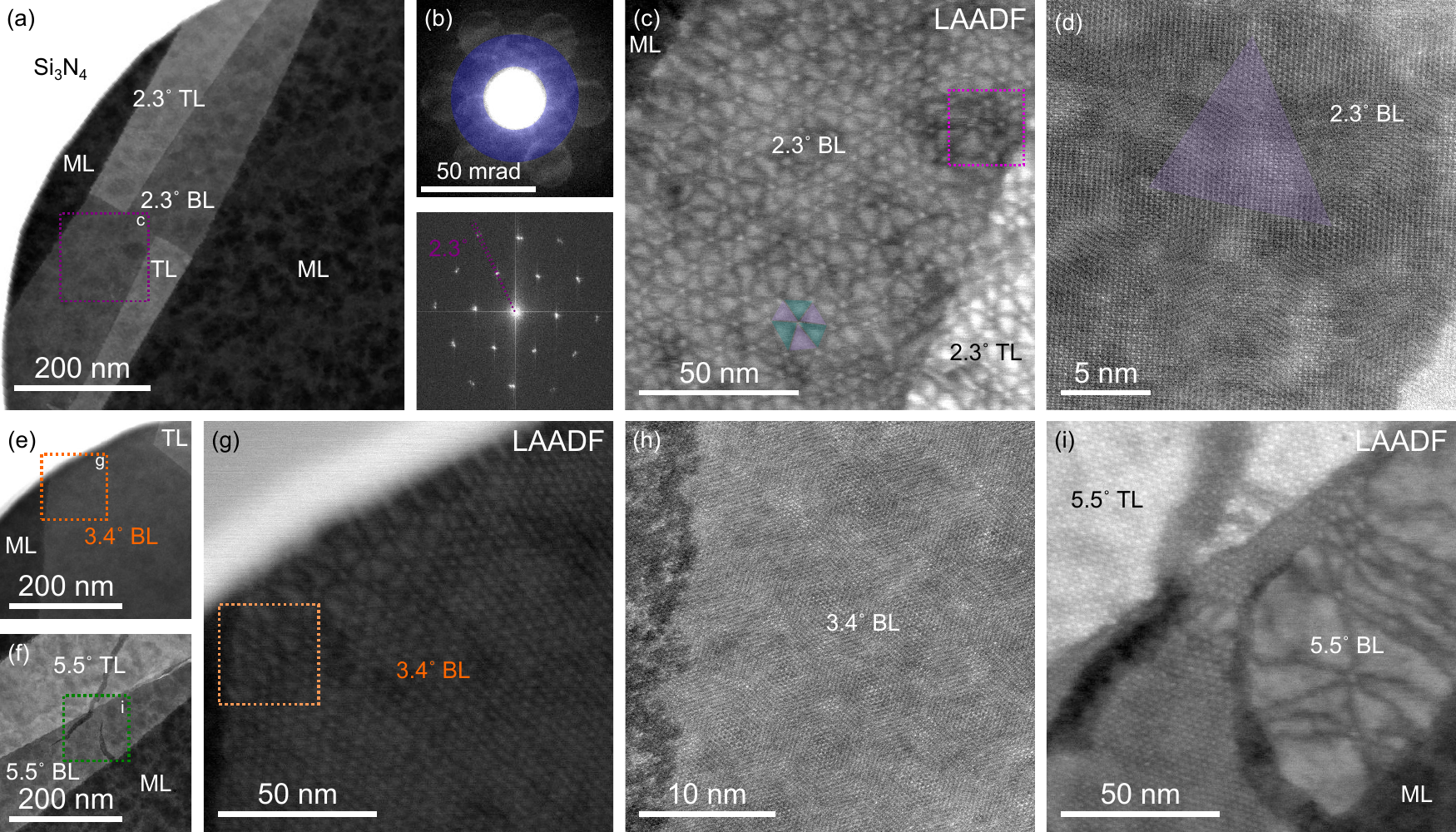}
\caption{\label{atomic-reconstruction}Low-magnification STEM-HAADF images of WSe$_{2}$ containing BLs and TLs with (a) 2.3$^{\circ}$, (e) 3.4$^{\circ}$ and (f) 5.5$^{\circ}$ twist angle. (b) Convergence and collection angle geometry for the STEM-LAADF imaging overlaid over a WSe$_{2}$ ML diffraction pattern used to observe the contrast at the bridge (Br) separating the AB/BA domains in the moir\'{e} unit cell. STEM-LAADF images of the WSe$_{2}$ BLs with low twist angles of (c) 2.3$^{\circ}$, (g) 3.4$^{\circ}$ and (i) 5.5$^{\circ}$, showing arrays of six-fold triangular domains that indicate AA-type stacking and atomic-reconstruction. Corresponding STEM-HAADF images of WSe$_{2}$ BLs with (d) 2.3$^{\circ}$ and (h) 3.4$^{\circ}$ twist angle from the regions marked in the LAADF images in (c) and (g), respectively.}
\end{figure*}

\begin{figure*}
\includegraphics[width=16.25cm]{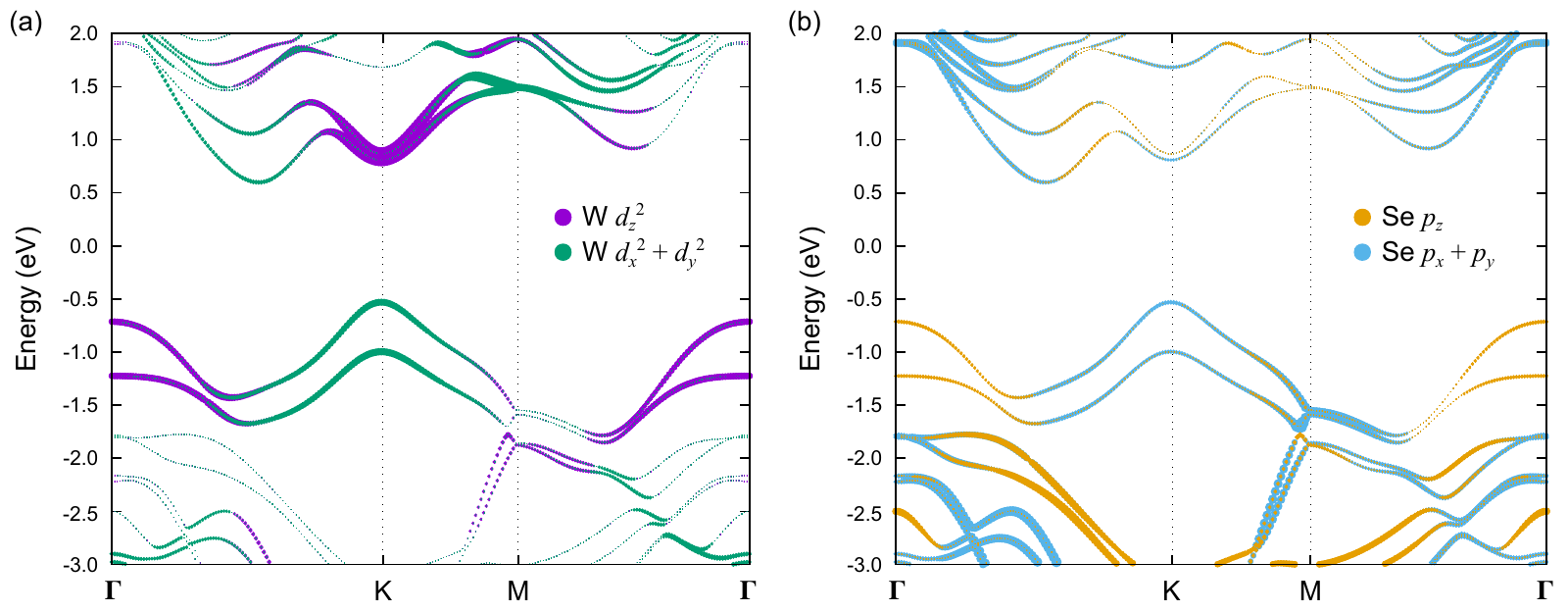}
\caption{\label{orbitals}Orbital-projected band structures of AA$^{\prime}$ stacked bilayer WSe$_2$ highlighting the orbital character separately for (a) the W atom $5d$ orbitals, and (b) the Se atom $4p$ orbitals. Symbol size is proportional to its orbital weight.}
\end{figure*}

\end{document}